
\input phyzzx
\def\lrpartial{\mathrel{\partial\kern-.75em\raise1.75ex\hbox
{$\leftrightarrow$}}} \line{ \hfill ULB-TH 03/95}
  \line{ \hfill January 1995}   \bigskip \centerline  {\bf OPERATOR WEAK
VALUES AND BLACK HOLE COMPLEMENTARITY\foot{presented at the Oskar Klein
Centenary Symposium (September 1994)}}  \bigskip
\centerline{F.ENGLERT\foot{E-mail: fenglert at ulb.ac.be}} \medskip
 \centerline {\it Service de Physique Th\'eorique}  \centerline{\it
Universit\'e Libre de Bruxelles, Campus Plaine, C.P.225 }
\centerline{\it Boulevard du Triomphe, B-1050 Bruxelles, Belgium}
 \centerline{and}
 \centerline {\it School of Physics and Astronomy}
 \centerline{\it Raymond and Beverly Sackler Faculty of Exact Sciences}
\centerline{\it Tel-Aviv University, Ramat-Aviv, 69978 Tel-Aviv, Israel}
\bigskip
\noindent {\bf Abstract.}
In conventional field theories, the
emission of Hawking radiation in the background of a collapsing star
requires transplanckian energy fluctuations. These fluctuations are
encoded in the weak values of the energy-momentum operator constructed
from matrix elements between both -in and -out states. It is argued that
taming of these weak values by back-reaction may lead to geometrical
backgrounds which are also build from weak values of the gravitational
field operators. This leads to different causal histories of the black
hole as reconstructed by observers  crossing the horizon at different
times but reduces, in accordance with the equivalence principle, to the
classical description of the collapse for the proper history of the star
as recorded by an observer comoving with it.  For observers never
crossing the horizon, the  evaporation would be interpreted within a
topologically trivial ``achronon geometry"  void of horizon and
singularity: after the initial ignition of the radiation  from pair
creation out of the vacuum of the collapsing star of mass M, as in the
conventional theory, the source of the thermal  radiation would shift
gradually to the star itself in a time at least of order $4M\ln 2M$. The
burning of the star could be consistent with a quantum unitary evolution
along the lines suggested by 't Hooft. A provisional formal expression of
 general  black hole complementarity is proposed and its possible
relevance for testing features of a theory of quantum gravity is
suggested.
\vfill\eject

\noindent {\bf 1. General Considerations. } \medskip Black hole
evaporation questions acutely the consistency of quantum physics and
general relativity. In the original derivation of black hole
radiance$^{[1]}$, the thermal density matrix describing the radiation
comes out  as a consequence of tracing pure quantum states over  states
hidden from an external observer by the event horizon. Disappearance of
the black hole would then result in a violation of unitarity within the
universe of external observers, as  suggested by Hawking$^{[2]}$.  A
halt of the  evaporation process at the Planck mass could formally save
unitarity by correlating the distant radiation to a stable planckian
remnant$^{[3]}$. The latter could then be generated in a  huge and
probably  infinite number of ways.  Thus, the end point of the
evaporation process would leave us with the problem either to understand
why, despite its breakdown  at the Planck scale, quantum physics is
operative at larger scales, or to enlarge in a consistent way the
framework of ordinary quantum field theory to incorporate infinitely
degenerate Hilbert spaces to describe this Planck scale.

An alternative approach to the dilemma has been proposed by 't
Hooft$^{[4],[5]}$. The Hawking radiation would induce some strong
back-reaction on the geometry which would appear, to the external
observer, free of singularity and horizons. The emission process would
be   part and parcel of a fully unitary evolution of the   black hole.
Susskind and al$^{[6]}$ have suggested to implement this idea by
materializing in the planckian vicinity of the event horizon a physical
``streched horizon", invisible to the free falling observer, where the
incoming information is deposited and then burned away. In this way
full evaporation would   be consistent with unitarity resulting from
long time correlations between otherwise thermal quanta.  As the very
existence of the  streched horizon would be observer dependent, they
argue that   unitary evolution  implies a departure from hitherto
admitted physical realism as expressed by the concept of an absolute,
observer independent, event. They characterize  such a relativity of
events as a kind of ``complementarity".

Unitarity is not the only conceptual  problem posed by black hole
evaporation. In conventional quantum field theory     the
emergence of   black hole radiance from  vacuum fluctuations in the
background of a collapsing star requires, because of the unbounded
blueshift in the horizons vicinity, frequencies well above the Planck
scale. In particular, any typical Hawking quantum detected at future
light-like infinity  $\cal{I}^+$ originates at past light-like
infinity   $\cal{I}^-$ from transplanckian frequencies highly tuned on
distances much smaller than the Planck size and these transplanckian
fluctuations constitute a gauge invariant effect$^{[7],[8]}$.
Transplanckian fluctuations  on  $\cal{I}^-$ do not pose any problem as
long as gravitational non-linearities are not taken into account, but
one would expect that their detailed structure should not lead to
observable effects on cisplanckian scales.  Nevertheless any smearing of
these fluctuations above the Planck scale would totally invalidate the
conventional derivation of the Hawking radiation.

There is however little doubt that Hawking radiation does occur because
it appears to follow from general thermodynamic considerations. The
information stored in the event horizon gives credence to   the
Bekenstein conjecture$^{[9]}$ that the area of the event horizon is a
measure of entropy. This entropy must then be, for dimensional reasons,
inversely proportional to the Planck constant and this in turn  requires
that static matter surrounding  an eternal black hole would thermalize
at a global temperature  proportional to $\hbar$. Consider indeed the
classical Killing identity$^{[10]}$ for a black hole surrounded by
static matter. This identity may be viewed  as the integrated constraint
equation over a static coordinate patch and  can be written as$^{[11]}$
 $$ \delta M_\infty= \delta
H_{matter} + {\kappa \over 2\pi} \delta {A\over 4}  \eqno(1)$$
  where $\kappa$ is the surface gravity of the hole, $M_\infty$ the
total mass at infinity   and $\delta H_{matter}$ is the variation of all
non gravitational parameters in the matter hamiltonian outside the
horizon. In the thermodynamic limit $\delta H_{matter}$ is the variation
of all interactions at the microscopic level and can be written as $T
dS_{matter} + \sum_i \mu_i d N_i$ where $S_{matter}$ is the matter
entropy and the sum is over all matter constituents. Thus, the
Bekenstein conjecture also implies that there exists a global
equilibrium    temperature $T$ proportional to the surface gravity for a
system composed of black hole and matter. But Eq.(1) being a classical
equation, this temperature should be proportional to $\hbar$ in order
to cancel the $\hbar^{-1}$ in the entropy. This reasoning is
corroborated by the estimate of this temperature via euclidean
continuation of the metric, either for Green's functions$^{[12]}$, for
partition functions$^{[13]}$ or for tunneling amplitudes$^{[14]}$,
because euclidean continuation always leads to the required dependence
of the temperature on $\hbar$. Moreover these methods all yield the same
result, namely $$T={1\over 8\pi M}\eqno(2)$$ for a black hole of mass
$M$ in absence of surrounding matter\foot{Note however that there would
be a decrease by a factor of two in the Euclidian periodicity if both
asymptotically flat spaces connected by the black hole throat were
identified. This is the ambiguity by a factor of two in the temperature
raised by 't Hooft$^{[15]}$}. The same value was also originally
obtained for the temperature of the thermal radiation emitted by  an
incipient  black hole using the dynamics of free fields in the curved
background of a collapsing star$^{[1]}$.

The above thermodynamic argument suggests that, possibly up to a
multiplicative constant, the result Eq.(2) does not really rely on the
mechanism used to derive it but rather on its internal consistency, and
the original derivation of the Hawking radiation from a collapsing star
is   perfectly consistent in absence of gravitational back-reaction. The
fact that the  thermodynamic argument refers more directly to
hypothetical eternal black holes than to incipient ones does not weaken
its significance. In an incipient black hole the drop of the centrifugal
barrier close to the horizon results there in a thermalization of
predominantly high angular momentum quanta at the high local
temperatures consistent with the global Hawking temperature Eq.(2).
This  means that, close to its  horizon, incipient black  holes tend to
behave as eternal ones. Therefore thermodynamics indeed strongly
suggests that Hawking radiation  is  a necessary consequence of
collapse.

The emergence of transplanckian frequencies is not specific to
black-holes; it is rather a general feature of  event horizons. There is
therefore not {\it a priori\/} a relation between the transplanckian
problem and the black-hole unitarity problem. However both problems most
likely require some understanding of the nature of space-time at the
Planck scale and hence one may hope that solving the former problem may
shed light on the latter. In this paper we shall motivate and present a
conceptual framework in which the taming of transplanckian frequencies
can be envisaged. It will be seen that this framework accommodates the
approach to the unitarity issue proposed by
 't Hooft. It leads to an enlarged notion of complementarity which can
be formulated within the realms of quantum theory using the concept of
``weak value" introduced by Aharonov and al.$^{[16]}$. Clearly however,
in absence of any reliable dynamics at the Planck scale, our approach
will be qualitative and provisional and the concepts developed here may
well turn out to be wrong. But if they contain some elements of truth
they may be of some help in unravelling the underlying structure of
quantum gravity.

Let us now summarize our approach.

It is reasonable to suppose that transplanckian frequencies are tamed
to planckian ones by some hitherto unknown  Planck scale physics. The
thermodynamic argument suggests that detection of Hawking radiation
survives the taming process although the dynamics of the thermal
emission might be drastically altered. Taming would arise if
back-reaction would slowdown  the collapse  to a halt so  as to prevent
the blueshifted fluctuations to reach  the Planck scale at past
light-like infinity. We shall argue, on the basis of an analysis of the
energy-momentum tensor of the radiating field  in the background of a
collapsing star, that such a taming mechanism can be expected when  the
back-reaction of the radiation on the geometry is taken into account.
Energy and momentum of the collapsing star would then be converted into
heat by degrees of freedom at the Planck scale, giving rise to thermal
emission at the Hawking temperature. The star would then burn and
evaporate within a trivial geometrical background displaying neither
horizon nor singularity.  This mechanism for removing   transplanckian
frequencies is in line with     the original proposal of t' Hooft:
Hawking radiation would   remain in causal contact with the collapsing
matter and evaporation can therefore possibly be consistent with
unitarity. It is     at odds with the conventional description of black
hole evaporation where the information is entirely contained in the
collapsing star which causally separates from the emission process,
leaving a genuine thermal radiation void of any information.

For this scenario to make sense it will be necessary to assume that, at
a more fundamental level, the classical description of the collapsing
objects   results   from an average over quantum spreads of  planckian
constituents. Even so, it   apparently contradicts the equivalence
principle as the small classical gravitational field at the horizon
appears inconsistent with the huge acceleration required to bring the
collapse to a halt. This clash can be resolved when the metric field
$g_{\mu\nu}(x)$ is promoted to a   quantum operator. Indeed, the
above-mentioned analysis of the energy-momentum tensor will  suggest
that,  in presence of horizons, the geometrical background
$g_{\mu\nu}(x)$ in which matter evolution can be described should   not
be identified  with the expectation value of the corresponding
Heisenberg  operator $\hat g_{\mu\nu}(x)$ in the normalized ``in"
quantum state $\ket i $ and thus  $$ g_{\mu\nu}(x) \neq  \bra i \hat
g_{\mu\nu}(x) \ket i. \eqno(3) $$ Rather, the background geometry should
be determined, as anticipated by 't Hooft$^{[5]}$ from both   the ``in"
state $\ket i $   and the ``out" state. This will lead to different
causal histories of the black hole as reconstructed by observers
crossing the horizon at different times but will reduce, in accordance
with the equivalence principle, to the classical description of the
collapse for the proper history of the star as recorded by an observer
comoving with it.

To understand this point in qualitative terms, consider a detector
sensitive only to cisplanckian effects. We call such a detector an
observer.  Let us first confine the motion of this observer within the
space-time outside the event horizon of a sufficiently massive
collapsing star. It  will necessarily encounter radiation. The
radiation recorded by such   observers can be encoded in some
``out"-state . Thus  in the space-time available to ``external"
observers, there exist  outgoing states describing  a particular set of
detectable  quanta covering the whole history of the evaporating black
hole.  This information  about a particular decay mode can be added to
the characterization of the system by the Schroedinger state of the
star  before collapse, or equivalently by the corresponding Heisenberg
state $\ket i$. More precisely we could specify that the system is
likely to be  be found at sufficiently late times in a state
characterized by some typical distribution of Hawking quanta. It may
seem at first sight that this added information about the future
detection of Hawking radiation is irrelevant for the   analysis of the
energy momentum tensor and of the metric at intermediate times. This
could be an incorrect conclusion for reasons we now explain.

The expectation value $ \bra i \hat A(t) \ket i$  of a Heisenberg
operator $\hat A(t)$ in the normalized  quantum state $\ket i $ is often
expressed as    $$ \bra i \hat A(t) \ket i =\sum_\alpha
P_\alpha\,A_\alpha \qquad   P_\alpha =\vert\langle \alpha  \ket i\vert^2
\eqno(4) $$ where the eigenvectors $\ket \alpha$ relative to eigenvalues
$A_\alpha$ form a complete set of orthonormal states. One   interprets
then  $  \bra i \hat A(t) \ket i$  as the average over the probability
distribution $P_\alpha$ of finding the value $A_\alpha$ if exact
measurements of a   complete set of commuting observables containing
$\hat A(t)$  are performed at time $t$ on a quantum system
``pre-selected" to be in the initial Shroedinger state $\ket{t_1,i} =
U(t_1,t_0)\ket i$ at time $t_1$. $U(t_1,t_0)$ is the evolution operator
to the time  $t_1$ from the time $t_0$ where the Shroedinger state is
identified with the Heisenberg one.

More information can   be gained if the system is also
``post-selected" to be found at a later time $t_2$ in a given
Shroedinger state $\ket{t_2,f}$$^{[16]}$.  One may then express
expectation values  $\bra i \hat A(t) \ket i$ as an average of   weak
values defined for $ t_1<t<t_2$ by $$A^{weak}_f  \equiv {\bra f \hat
A(t) \ket i \over \bra f i\rangle}\eqno(5)$$  where $\ket f = U(t_0,t_2)
\ket{t_2,f}$.  One gets   $$ \bra i \hat A(t) \ket i = \sum_f
P_f\,A^{weak}_f \qquad   P_f =\vert\langle f \ket i\vert^2.\eqno(6) $$
Eq.(6) suggests that weak values represent   measurable quantities for a
pre- and post-selected system. This is indeed the case if a measurement
of $\hat A(t)$ is performed on the system with sufficient quantum
uncertainty to avoid disrupting the evolution of the system. Such
``weak" measurements yield not only the real part of  $A^{weak}_f$ but
also its imaginary part and reconstruct in this way the available
history between $t_1$ and $t_2$ for a system  pre-selected at $t_1$ and
post-selected at $t_2$$^{[16]}$.

Generally, the information gained by post-selection and weak
values is relevant only if one post-selects a state   describing   rare
events for the pre-selected state considered, that is if $P_f$ is
located in the tail of the distribution probability. However, the
situation is different when one considers the Hawking emission process in
the classical background of a collapsing star. There, in absence of
back-reaction, the energy-momentum tensor of the radiation can be
computed exactly in some simplified  models. It then appears
that post-selected states defined on a space-like surface $\cal \sigma$
arbitrarily close to the union of the event horizon  $\cal H$ and of the
future light-like infinity $\cal {I}^+$ may yield weak values of the
energy-momentum tensor operator $\hat T_{\mu\nu}(x)$ very different
from its average value. While the latter remains  smooth on the scale of
the Schwartzschild   radius, the former may  exhibit in that region
oscillations of unbounded amplitudes and  a singularity on the event
horizon$^{[7]}$. These features, which are a  consequence of the
unbounded blueshift experienced in the vicinity of the horizon by the
vacuum fluctuations generating the Hawking quanta, persist in
generic post-selected states detectable by external observers.  The
energy content of these  fluctuations show up in the weak values of
$\hat T_{\mu\nu}(x)$ but are averaged out in expectation values. However
observers who do cross the horizon detect different post-selected
states. These yield weak values of  $\hat T_{\mu\nu}(x)$  which
are  smooth as the observer approaches the horizon.

 Taming the energy fluctuations   will lead rather
naturally to a back-reaction picture in which the geometry
reconstructed by external observers exhibits neither horizon nor
singularity. This geometry would not be determined by the expectation
value of the metric but would be obtained from the weak values
 $$g_{\mu\nu}(x)^{weak}_f \equiv {\bra f \hat g_{\mu\nu}(x) \ket i \over
\bra f  i\rangle}.\eqno(7)$$
 More precisely, expressing as in Eq.(6) expectation values in terms of
weak values  $$ \bra i \hat g_{\mu\nu}(x) \ket i = \sum_f P_f {\bra f
\hat g_{\mu\nu}(x) \ket i \over \bra f  i\rangle}\qquad P_f
=\vert\langle f  \ket i\vert^2 \eqno(8)$$ we  are lead to  the
assumption that, because of the sensitivity of the horizon to
back-reaction  effects the right hand side is not dominated by a single
classical geometry but by a set of distinct, possibly complex,
geometries depending on the post-selected state. A distinct geometrical
background would then be defined by a restricted average
$<g_{\mu\nu}(x)>_{\{\alpha\}}$ over weak values for post-selected states
$f_{\{\alpha\} }$ in Eq.(8) resulting from keeping
fixed some macroscopic parameters $\{\alpha\}$, namely   $$
<g_{\mu\nu}(x)>_{\{\alpha\}}= \sum_{f_{\{\alpha\}\epsilon \{\alpha \} }}
P_{f_{\{\alpha\}}}  {\bra {f_{\{\alpha\}}}  g_{\mu\nu}(x) \ket i \over
\bra {f_{\{\alpha\}}} i\rangle}; \qquad P_{f_{\{\alpha\}}}\equiv
{\vert\bra {f_{\{\alpha\}}} i\rangle\vert^2 \over \sum_{
f'_{\{ \alpha'\}}} \vert\bra { f'_{\{ \alpha'\}}}
i\rangle\vert^2}\eqno(9)$$

In this way the the fully evaporated black hole  in tamed vacuum would
be characterized by all the post-selected  Hawking radiation states
$\ket f$ on $\cal {I}^+$ which would then constitute a Cauchy surface in
the future of $x$; the resulting geometry   connecting   $\cal{I}^-$  to
$\cal{I}^+$ would be free from the black hole horizon and  singularity
and would serve as a background for reconstructing its history.
Geometries corresponding to no detectable outgoing Hawking quanta would
require a different future Cauchy surface and a different
post-selection. They would uncover the classically singular  background
of the collapsing star with its original mass $M$, and  the proper
history recorded by the comoving observer which detects essentially no
radiation would then  agree with the classical predictions in accordance
with the equivalence principle. Post-selections corresponding to
observers in free fall after recording a partial evaporation   would
generate a geometrical background  accommodating a black hole with
mass $m$ smaller than $M$. The role of $\{\alpha\}$ is played here by
the final mass $0\leq m \leq M$ of the star and by the geometry of the
future Cauchy surface which should be self-consistently dynamically
determined from the final mass $m$.

In the following sections we shall discuss these ideas within
simplified models by confronting the problems met in the conventional
derivation of the Hawking radiation based on local field theory. In
section 2, we analyse, in the vicinity of the horizon,  the  discrepancy
between weak and  expectation values  of  the energy-momentum tensor in
the classical background of a collapsing star. This analysis is based on
the work of Massar and Parentani$^{[7]}$ who generalize the concept of
weak values. They introduce the notion of partial post-selection on a
subspace of the full Hilbert space to cope with the hiding of
information by event horizons. We shall present the considerations
relevant for our analysis in a comprehensive manner and we shall refer
the reader interested in further development on partial post-selection
to reference $[7]$.
 These result are used in section 3 to motivate the proposed taming
mechanism and its obvious implications for the unitarity issue. We
discuss its consistency with Lorentz invariance and causality. The
general formal expression Eq.(9) is proposed as a provisional expression
of black hole complementarity and its possible relevance for uncovering
features of a theory of quantum gravity is suggested.
 \bigskip \noindent
{\bf 2. Weak and Expectation Values in the Background of a
Collapsing         Shell. } \medskip
 Let us first review the Hawking emission process for a classical
collapsing
 spherically symmetric shell of mass $M$.  Instead of specifying the
energy momentum of the shell and deducing from it its trajectory, we
shall take, as in references $([17],[18],[8])$ the latter as the input
of the analysis. Inside the shell, Minkowski space can be described by
the coordinate system $$\eqalign{&ds^2= dU\ dV -r^2 d\Omega^2 \cr
&U=\tau - r \qquad V=\tau + r,}\eqno (10)    $$ for $r < r_s(\tau)$
where $r_s(\tau)$ is the shell radius at time $\tau$. The input
parameter will be taken to be $w_s$ , the radial inwards velocity of the
shell with respect to an observer at rest inside the star, namely
$$  w = -{d r_s(\tau) \over d\tau}.\eqno (11)   $$ Outside the shell and
outside the event horizon we parametrize  space-time  by tortoise
coordinates $$ \eqalign {  &ds^2 = g_{00}(r) du  dv - r^2 d\Omega^2
\qquad  g_{00}(r)\equiv \left(1- {2M\over r}\right) \cr &u= t-r^* \qquad
v=t+r^*\cr &dr = (1- {2M\over r}) dr^* ,}\eqno (12)   $$ where $r$,
understood as a function of $v-u$, is the ``radius" which measures the
invariant surface $4\pi r^2$ of a sphere and is continuous across the
shell. The entire space-time  outside the event horizon  can be
parametrized by a single $(u,v)$ coordinate system by performing  inside
the shell a coordinate transformation  $U(u), V(v)$. Continuity of
$ds^2$ and $r$ across the shell imply on the shell $$ \eqalignno
{{dU\over du}{dV\over dv}&=g_{00}(r_s)&(13)  \cr dV-dU
&=g_{00}(r_s)(dv-du)&(14)}$$ An outgoing photon emitted from inside the
star is redshifted on ${\cal I}^+$ at $v=+\infty$ by a factor $dU/du$.
 From Eqs (13) and (14), $$ {dU\over du} = \left[ - w +\sqrt
{w^2+(1-w^2)g_{00}(r_s)}\right] {1\over 1-w} \eqno(15)$$ where $r_s$ is
the radius of the shell intersected by the outgoing ray at the value of
$u$ considered. For a shell at rest, one gets $$ {dU\over du} =
g_{00}^{1/2}(r_s). \eqno(16)$$
 For a collapsing star, in the vicinity of the horizon where $
g_{00}^{1/2}(r_s) \to 0$ and $w \to \alpha$ where $\alpha$ is a constant
such that $0< \alpha \leq 1$, we get the asymptotic value $$ {dU\over
du} =  g_{00}(r_s){1+\alpha \over 2\alpha}.\eqno(17)$$ Eqs.(16) and (17)
show clearly the enhanced redschift of a photon emitted from inside the
star due to the motion of the shell.

 Let us first examine the s-wave contribution from a free scalar field
to the radiation. Neglecting the residual potential barrier, the
Heisenberg scalar field operator rescaled by $r$ obeys then $$
\partial_u \partial_v \Phi=0.\eqno(18)$$ It is expanded
 into a complete set of   solutions $\phi_k$, that is $$ \phi_k = f_k(u)
+ g_k(v)   \eqno (19) $$ such that $$ (\phi_j\vert \phi_i) \equiv
i\int_{\Sigma}
 [\phi_j^*\lrpartial_v \phi_i dv - \phi_j^*\lrpartial_u \phi_i du ] =
\delta (i-j) \eqno(20) $$ where $\Sigma$ is an arbitrary
 Cauchy surface which does not cross the horizon  and the $\phi_k$
 vanish at $r=0$.

Moving  back in time, positive  frequency plane waves solutions defined
on ${\cal I}^+$   reflect on a timelike  curve $V(v)-U(u) =2r=0$ and
propagate to ${\cal I}^-$ where they span only the domain $v<0$ (the
zero of $v$ is chosen here to coincide on ${\cal I}^-$ with the last
rays arriving on ${\cal I}^+$).  Hence wave-packets build out of positive
frequencies in the retarded time $u$ on ${\cal I}^+$ require frequencies
of both signs in the advanced time $v$ on ${\cal I}^-$. To build a
complete set of solutions of the field equations with positive
frequencies on ${\cal I}^-$  one needs waves   having support on
$v>0$  and propagating towards the horizon. The Heisenberg vacuum
$\ket{i}$ annihilated by destruction operators  associated with positive
frequencies waves on  ${\cal I}^-$ is then a superposition of pairs
formed from outgoing quanta on ${\cal I}^+$ correlated to partner states
described by waves propagating towards the horizon $\cal H$. While the
outgoing quanta are real particles, this is not true for the partners
quanta which are associated to waves with  no definite frequency sign
and should be interpreted as vacuum fluctuations. Taking a trace of
$\ket {i}\bra{i}$ over the latter states, one gets a density matrix
describing at sufficiently large retarded times $u$ a thermal flux of
outgoing particles  on ${\cal I}^+$ at the Hawking temperature Eq.(2).
For definiteness we exhibit here the various waves in the case of a
light-like shell $(w=1)$. Denoting respectively by $\vert
-\omega^{out})$ and  $\vert +\omega^{out})$ the outgoing plane waves and
their partner we have$^{[8]}$ $$\eqalignno{&\vert -\omega^{out})
 ={1\over \sqrt{4\pi \omega}}  \left[ \exp (-i\omega u) -\Theta(-v) \exp
(i4M\omega
 \ln {-v\over A})\right]&(21) \cr &\vert +\omega^{out})={1\over
\sqrt{4\pi \omega}}\Theta(v) \exp (-i4M\omega \ln {v\over A})&(22)
\cr}$$ where  the frequencies $\omega$ span the
 positive real axis. A convenient complete set of positive frequency
in-modes on  $\cal {I}^-$   is, taking $u \to\infty$,   $$ \eqalign {&
\vert \pm\omega^{in})={1\over
 \sqrt{8\pi \omega\ \sinh (\omega 4\pi M)}} \Theta(v)    \exp (\mp
i4M\omega \ln {v\over A} ) \exp (\pm \omega 2\pi M) \cr &+{1\over
 \sqrt{8\pi \omega\ \sinh (\omega 4\pi M)}} \left[ \Theta(-v) \exp (\mp
i4M\omega \ln {-v\over A}) \exp (\mp \omega 2\pi M )- \exp(\pm i\omega
u) \right]  .}\eqno(23)$$ Defining the destruction operators
$a^{in}_{\pm\omega}$ and  $a^{out}_{\pm\omega}$ associated to the modes
Eqs.(21),(22) and (23), we have   $$\eqalignno{\Phi(u,v) =
 &\int_0^\infty  d\omega \left[\vert -\omega^{out}) a^{out}_{-\omega} +
h.c.\right] &\cr + &\int_0^\infty  d\omega \left[\vert +\omega^{out})
a^{out}_{+\omega} + h.c.\right].&(24)\cr} $$ or equivalently  $$
\Phi(u,v) = \int_0^\infty
 d\omega \left[\vert -\omega^{in}) a^{in}_{-\omega} + \vert
+\omega^{in}) a^{in}_{+\omega} + h.c.\right] .\eqno(25)$$
 The Bogoliubov transformation relating in- and out- operators take the
form
 $$ \eqalign {  a^{in}_{+\omega}
  &=  \alpha_\omega a^{out}_{+\omega} - \beta_\omega a^{out\
\dagger}_{-\omega} \cr a^{in}_{-\omega} &=   \alpha_\omega
a^{out}_{-\omega} - \beta_\omega a^{out\ \dagger}_{+\omega}}\eqno(26) $$
where $$ \alpha_\omega ={\exp
 (\omega 2\pi M)\over \sqrt{ 2\sinh (\omega 4\pi M)}} \quad \beta_\omega
={\exp (- \omega 2\pi M)\over \sqrt{ 2\sinh (\omega 4\pi
M)}}.\eqno(27)$$ From Eqs.(26) and (27) one easily verifies Eq.(2).

The transplanckian problem arises because Hawking quanta arriving at
late time $u_l$  on $\cal {I}^+$ are build out of  vacuum fluctuations
with very high frequencies on  $\cal {I}^-$ in a the Lorentz frame fixed
by the spherical symmetry of the shell.  We represent a typical quantum
detected  at the retarded time $u_l$ on ${\cal I}^+$ by a   wave-packet
centred, in accordance with Eq.(2) around a positive frequency
$\omega=O(1/2M)$ and extending over a distance of the order of its
wavelenght $2M$. Its complex frequency spectrum $f(\omega)$ has the form
 $$f(\omega) =\vert f(\omega)\vert \exp (-i\omega u_l) \eqno(28)$$ where
$\vert f(\omega)\vert$ is  spread over a range comparable to its central
frequency. When this wave packet moves back in time with the velocity of
light, it   is blueshifted by $du/dU$ given in Eq.(17). In the vicinity
of the horizon, one may choose the arbitrary constant defining $r^*$ to
write, when $1-2M/r <1$,
 $$1-{2M\over r}  \simeq  \exp\left({v-u \over 4M}\right). \eqno (29)$$
Hence at late retarded times $$g_{00}(r_s) =\exp\left({v_\infty-u \over
4M}\right), \eqno(30)$$ where $v_\infty$ is the asymptotic coordinate of
the collapsing shell on the horizon.  Labeling by $\tilde \omega$ the
central frequency of the packet traced back to ${\cal I}^-$  one thus
gets  from Eqs.(17) and (30)   $$  {\tilde\omega \over \omega}\simeq
g_{00}^{-1}(r_s) \simeq {-4M \over U(u_l)}\eqno(31)$$ where the zero of
$U$ is chosen on the horizon.  One may verify Eq.(31) from the explicit
expression of the wave Eq.(23) which yields a local frequency $\tilde
\omega = 4M\omega /  (- v)$; for $w=1$   $v=V$ and the relevant
waves satisfy $U(u_l) - V =U(u_l) - v=0$.  The increase in frequency
from $\omega$  to $\tilde \omega$  is accompanied by   a localisation
within a wavelenght $\tilde\omega^{-1}$. The wave-packet propagating
backwards in time reflects on $r=0$ and is correlated on ${\cal I}^-$ to
a partner centred at  $v>0$. This correlation arises because a plane
wave of frequency $\tilde\omega$  extends on both sides of $v=0$. The
partner with $v>0$ can then be similarly depicted as a wave packet with
the same frequency $\tilde\omega$ as the ancestor of the Hawking quantum
but the separation between the two packets is on ${\cal I}^-$ comparable
to their spread$^{[7],[8]}$.

Although in absence of back-reaction the Planck scale does not enter the
problem, it is useful for the forthcoming discussion of section 3, to
characterize the different time scales for Hawking emission in the
space-time background used here  with respect to this scale. We first
note that Hawking emission starts  about  a  retarded time  $u_0$ (and a
corresponding Schwartzschild  time $t_0$) when the asymptotic value
Eq.(30) becomes accurate, that is when   $v$ on
the star surface can be well approximated by the constant $v_\infty$.
This happens when $\delta\equiv r-2M$ is still of order $2M$ and we
label the corresponding coordinate separation by $\delta_0$. Thus  $$
\delta_0 =O(2M). \eqno(32)$$
 A typical Hawking quantum requires, from Eqs.(30) and (31)
transplanckian vacuum frequencies
 $\tilde \omega >1$ inside the shell and on ${\cal I}^-$ after  retarded and
Schwartzschild times $u_1$ and $t_1$ such that $$t_1 - t_0 = {u_1 -u_0
\over 2} =O (2M \ln 2M). \eqno (33) $$ while for a static observer just
outside the shell the frequencies remain cisplanckian up to times such
that $\omega / \sqrt g_{00}(r_s) \simeq 1/ 2M \sqrt g_{00}(r_s)$ becomes
of order unity, that is $$t_2 - t_0 = {u_2 -u_0 \over 2} =O (4M \ln 2M).
\eqno (34) $$ The corresponding location of the centre of the packets in
the vicinity of the shell are, from Eq. (30) $$ \delta_1 = O(1);\qquad
\delta_2 = O(1/2M).\eqno(35)$$ Note that, at fixed time $t$, $\delta_1$
measures a planckian   distance to the horizon inside the star while
$\delta_2$   measures  a planckian distance to the horizon  outside the
star, as follows from the metric Eq.(12). At time $t_2$, the local
temperature   $T/ g_{00}^{(1/2)}(r_s)$ reaches, from Eq.(35), the Planck
temperature just outside the shell.

   The transplanckian vacuum fluctuations give rise to corresponding
transplanckian weak-values of the energy-momentum tensor for
post-selected states describing Hawking quanta and these unbounded
fluctuations are responsible for the discrepancy between average and
weak values in the vicinity of the horizon. In general, post-selected
states are defined by eigenstates of a complete set of commuting
observables on some Cauchy surface $\Sigma_f$ laying in the future of
the space-time points $\{x\}$ where weak values are computed. We shall
take for $\Sigma_f$ a space-like surface infinitesimally close to the
union of the future light-like infinity ${\cal I}^+$ and of the event
horizon $\cal H$. We shall consider the subset of post-selected states
which can be obtained by measuring devices sensitive only to
cisplanckian frequencies.  This can be achieved in the following way. We
write the full Hilbert space $H$ of post-selected states as $$ H = H_{
\cal H} + H_{{\cal I}^+}\eqno(36)$$ where $ H_{ \cal H} $ and $ H_{{\cal
I}^+}$ are engendered respectively by the operators acting in  $\cal H$
and in  ${\cal I}^+$. Consider a post-selection which would be defined
only in the Hilbert space $ H_{{\cal I}^+}$ and let $\ket {P_{1\alpha}}$
be such a   state. This state is EPR correlated to a state  $\ket
{P_{2\beta}}$ which is, up to a normalization constant, equal to $\bra
{P_{1\alpha}} i\rangle$. The corresponding post-selected state is,
up to normalization, $\ket {P_{1\alpha}} \bra {P_{1\alpha}} i\rangle$
and we shall call it a partially post-selected state. Thus  the weak
value of the renormalized energy-momentum tensor operator  $\hat
T_{\mu\nu}$ of the scalar field $\Phi$, corresponding to this partially
post-selected state,  is  according to Eq.(5)
$$T_{\mu\nu}(x)^{weak}_{P_{1\alpha}}  \equiv {\bra i P_{1\alpha}
\rangle \bra {P_{1\alpha}} \hat T_{\mu\nu}(x) \ket i \over \bra i
P_{1\alpha} \rangle \bra {P_{1\alpha}}  i\rangle}. \eqno(37)$$ Note that
one may insert in Eq.(37) the projection operator $I_{ H_{ \cal H}}$ on
the Hilbert space  $H_{ \cal H}$ and define a post-selection density
matrix $\Pi_S \equiv  I_{ H_{ \cal H}}\ket {P_{1\alpha}}\bra
{P_{1_\alpha}}$. 	In this way,  Eq.(37) can be rewritten in the
formulation of partial post-selection given in reference [7], namely  $$
T_{\mu\nu}(x)^{weak}_{P_{1\alpha}}\equiv { \bra i  \Pi_S \hat
T_{\mu\nu}(x) \ket i  \over \bra i  \Pi_S \ket i  }. \eqno(38)$$ The
probability of finding $\ket {P_{1\alpha}}$ by tracing over  $ H_{ \cal
H} $ is $$ p_\alpha =  \bra i P_{1\alpha} \rangle \bra {P_{1\alpha}}
i\rangle \eqno(39)$$ and it is easy to check that it is equal to the
probability of finding simultaneously
 $\ket {P_{1\alpha}}$ and its EPR-correlated state in $\ket i$. We have
$$ \sum_\alpha p_\alpha  T_{\mu\nu}(x)^{weak}_{P_{1\alpha}} =
\sum_\alpha \bra i I_{ H_{ \cal {I}^+}} \hat T_{\mu\nu}(x) \ket i = \bra
i \hat T_{\mu\nu}(x) \ket i .\eqno(40)$$ Eq.(40) expresses  that summing
weak values over all partially post-selected states  on ${\cal I}^+$
yields back again the expectation value of $  \hat T_{\mu\nu}(x)$. Each
of the partially post-selected states  $\ket {P_{1\alpha}}$ can be
interpreted as a state which could have been detected by observers whose
motion is confined to the space-time region bounded by the event
horizon. We shall say that these are the post-selected states available
to external observers: they contain the maximal amount of information
accessible to these observers that can be added to the one contained in
the state $\ket i$.

 To compare the weak values Eq.(37) with the expectation value of  $\hat
T_{\mu\nu}$ we  write
 $$ \hat T_{\mu\nu}(x) = :\hat T_{\mu\nu}(x): + \, T_{\mu\nu}(x) \eqno
(41)$$ where $:\hat T_{\mu\nu}(x):$ stands for the normal ordered
operator  in the Heisenberg state $\ket i$; hence the c-number
$T_{\mu\nu}(x)$ is $$T_{\mu\nu}(x)=\bra i \hat T_{\mu\nu}(x) \ket i
.\eqno(42)$$

Outside the star, $T_{\mu\nu}(x)$ can be computed exactly in the
2-dimensional approximation corresponding to the reduced field equation
Eq.(18). In that approximation the 2-dimensional energy-momentum tensor
operator $\hat T_{\mu\nu}^{(2)}(x) = \hat T_{\mu\nu}(x) 4\pi r^2$. One
has $$ :\hat T_{\mu\nu}^{(2)}(x): =: \partial_\mu \Phi  \partial_\nu
\Phi-{1\over2}g_{\mu\nu} (g^{\sigma\tau} \partial_\sigma \Phi
\partial_\tau \Phi):. \eqno(43)$$ The expectation value of $\hat
T_{\mu\nu}^{(2)}(x)$ follows$^{[19]}$ from the trace anomaly $$
T_{uv}^{(2)}(x)= -{1\over 12\pi} \partial_u \partial_v \rho, \eqno(44)$$
where $\exp (2\rho)$ is the conformal factor in the $(u,v)$ coordinate
system which coincides with $g_{00}(r)$ outside the star, and from the
conservation law $$ \hat T^{\mu (2)} _{\nu;\mu}(x)  = 0. \eqno (45)$$
One gets  outside the shell
 $$\eqalignno {4\pi r^2 T_{uu}(x)=\bra i \hat T_{uu}^{(2)}(x) \ket i
&= {1\over 12\pi} \left[ -{M\over 2r^3}(1-{2M\over r}) - {M^2\over
4r^4}  \right] + t_u(u)  &(46)\cr 4\pi r^2 T_{vv}(x) =\bra i \hat
T_{vv}^{(2)}(x)  \ket i    &= {1\over 12\pi}  \left[ -{M\over
2r^3}(1-{2M\over r}) - {M^2\over 4r^4} \right] + t_v(v)  &(47)\cr  4\pi
r^2 T_{uv}(x) =\bra i \hat T_{uv}^{(2)}(x)  \ket i   &= -{1\over 12\pi}
{M\over 2r^3}(1-{2M\over r}). &(48)}$$
 The last terms in Eqs.(46) and (47) are determined from  boundary
conditions. The vanishing of $ T_{\mu\nu} $ on $\cal {I}^-$ gives
$t_v=0$ and the regularity of $ T_{\mu\nu} $ on the horizon in an
inertial frame requires that $T_{uu}$ vanishes when $u\to \infty$ as
$O(1 - 2M/r)^2$ or equivalently  $$ t_u = {\pi \over 12}{ 1\over (8\pi
M)^2}. \eqno (49)$$ This term represents from Eq.(46) the outgoing
energy flux as $r\to \infty$ and indeed coincides with the thermal flux
of the Hawking radiation.

These values of $t_u$ and $t_v$ characterize the boundary conditions of
the ``Unruh vacuum" in contradistinction with the values $t_u =t_v=0$.
The latter define the ``Boulware vacuum" where  $ T_{\mu\nu} $ vanishes
both on $\cal {I}^-$ and $\cal {I}^+$. As follows from Eq.(46) the
energy density  then diverges negatively in an inertial frame as one
approaches the horizon. The Boulware vacuum  mimics the vacuum energy
that would be provoked by a static shell sitting on the horizon.

 On the other hand, we  may write, using the Bogoliubov transformations
 $:\hat T_{\mu\nu}(x):$   acting on $\ket i$ as $$:\hat
T_{\mu\nu}(x):\ket i = ::\hat T_{\mu\nu}(x)::\ket i + \Delta
T_{\mu\nu}(x)\ket i \eqno(50)$$ where $::\hat T_{\mu\nu}(x)::$ contains
only creation operators on out-states and $\Delta T_{\mu\nu}(x)$ is a
c-number. Thus
 $$\Delta T_{\mu\nu}(x)= {\bra {f_0}: \hat T_{\mu\nu}(x): \ket i \over
\bra {f_0}  i\rangle}. \eqno(51)$$ where $\ket {f_0}$ labels the
out-vacuum.  The weak value Eq.(37) of the energy- momentum tensor for
the partially post-selected state $\ket{ P_{1\alpha}}$ can be written as
$$T_{\mu\nu}(x)^{weak}_{ P_{1\alpha}}  = { \bra i P_{1\alpha} \rangle
\bra {P_{1\alpha}}:: \hat T_{\mu\nu}(x):: \ket i \over  \bra i
P_{1\alpha} \rangle \bra {P_{1\alpha}}  i\rangle}+ \Delta T_{\mu\nu}(x)
+   T_{\mu\nu}(x)  . \eqno(52)$$
 Computing $\Delta T_{\mu\nu}(x)$ on $\cal {I}^+$ in the asymptotic
region $u>u_0$ , we get from Eqs.(43), (25) and (23)\foot{We use here
the waves $\vert \pm \omega_{in})$ for the light-like collapse $w=1$. It
is easily verified that in the asymptotic region the results
Eqs.(54),(55) and (56) are valid for the general collapse $0<w \leq
1$.}, using the Bogoliubov transformation Eq.(26),  $$  4\pi r^2 \Delta
T_{\mu\nu} =  - \int_0^\infty d\omega  {2\beta(\omega)\over
\alpha(\omega)} \partial_\mu \vert-\omega^{in})^*
 \partial_\nu \vert-\omega^{in}) \eqno(53) $$ or explicitely
$$\eqalignno { 4\pi r^2 \Delta T_{uu}&= -{1\over 2\pi} \int_0^\infty
d\omega { \omega \over e^{\omega 8\pi M } -1} = - {\pi \over 12}{1\over
(8\pi M)^2} &(54)\cr
 4\pi r^2 \Delta T_{vv}&= -{1\over 2\pi}{1\over (8\pi M)^2}{16M^2 \over
(v-i\epsilon)^2} &(55)\cr
 4\pi r^2 \Delta T_{uv}&=0. &(56)}$$ Going to $\cal {I}^+$ we get from
Eqs.(40) and (52) $$ \lim_{v\to +\infty} \sum_\alpha p_\alpha{ \bra i
P_{1\alpha} \rangle \bra {P_{1\alpha}} ::\hat T_{\mu\nu}(x):: \ket i
\over  \bra i P_{1\alpha} \rangle \bra {P_{1\alpha}} i\rangle} =
{\pi\over 12} {1\over (8\pi M)^2} \delta_{\mu u} \delta_{\nu u} \eqno
(57)$$  where the sum extends over all partially post-selected states on
$\cal {I}^+$. This is consistent with the interpretation  that $\bra i
P_{1\alpha} \rangle \bra {P_{1\alpha}} ::\hat T_{\mu\nu}(x)::  \ket i /
\bra i P_{1\alpha} \rangle \bra {P_{1\alpha}}  i\rangle$ represents  the
energy density contained in the
 post-selected Hawking quanta forming the state $\ket {P_{1\alpha}}$:
Eq.(57) tells that the average energy density  carried by the post
selected Hawking quanta is equal to that of the Hawking flux on  $\cal
{I}^+$. More generally Eq.(52) describes these quanta as if they were
travelling on a vacuum whose energy density  is
  $T_{\mu\nu} + \Delta T_{\mu\nu} $. Outside the shell $v>4M$ and
$\Delta T_{vv}$ becomes negligible compared to $\Delta T_{uu}$ as $v$
increases. Comparing Eqs.(54), (55) and (56) to Eqs.(46), (47), (48) and
(49) we see that the term $t_u$ has cancelled out in the new vacuum and
that outside the shell the post-selected quanta travel thus on a vacuum
whose energy density is essentially the one  due to a Boulware vacuum
characterized by $t_u=t_v=0$.

Thus, by post-selection we switch from a description where the energy of
the Hawking radiation appears as a property of the Unruh vacuum to a
description in which excited particles building the Hawking flux travel
on a Boulware vacuum. The latter has a negative energy density which
diverges  on the horizon in an inertial frame. Moving backward in time
this divergence   reflects back on $r=0$ up to $\cal {I}^-$, as seen
from Eq.(55). In fact $\Delta T_{vv}$ is symmetric with respect to $v=0$
and the vanishing of the total minkowskian energy on $\cal {I}^-$ is
ensured from Eq.(55) by a positive energy singularity carried by the
last rays $^{[7]}$. The last rays may be viewed as causally  generating
the horizon and thus the Boulware vacuum energy density is in that sense
causally related to energy densities lay down on $\cal {I}^-$. The
positive energy singularity carried by the last rays accordingly
generates a   singularity on the horizon itself, and this constitutes a
general feature of the Boulware vacuum$^{[20]}$. The equivalence on the
average  between the description in terms of   excited quanta on top of
a Boulware vacuum causally generated by the singular last rays and the
description in terms of the Unruh vacuum, is ensured by Eq.(40). This
equivalence is however the result of a very delicate interference
process. Indeed, the weak energy-momentum carried by the post-selected
quanta (the first term in Eq.(52)) in the vicinity of the last rays and
of the horizon exhibits oscillations with divergent amplitudes and
frequencies$^{[7],[8]}$. This is a consequence of the unbounded
differential blueshift which distort
 the small oscillations inherent to  wave packets built on $\cal {I}^+$
from positive frequencies
 only. On the contrary, the causal Boulware vacuum energy diverges
monotonically in the vicinity of the last rays and of the event horizon.
Therefore, even for generic partially post-selected states on  $\cal
{I}^+$  expectation and weak values of $\hat T_{\mu\nu}(x)$ will behave
very differently   and the unbounded fluctuations of
$T_{\mu\nu}(x)^{weak}_{P_{1\alpha}}$ will survive such generic
post-selection.

The post-selected states considered above contain the maximal amount of
information available for external observers. However an observer
comoving with the star will detect practically no quanta and cannot gain
information from post-selection. In other words, the partial post
selected states
  available to  comoving observers is    $I_{ H_{ \cal {I}^+}} \ket i
=\ket i $ and the corresponding weak value   $$ T_{\mu\nu}(x)^{weak}  =
\bra i \hat T_{\mu\nu}(x)\ket i \eqno (58)$$ then simply coincides with
the expectation value. Observers who  cross the horizon at later advance
time  $v_c$ than the shell stop detecting radiation after some critical
retarded time $u_c(v_c)$. They
 can be similarly be characterised by some post-selected states
containing the maximal information available to them. We surmise that
these would lead, as in Eq.(58), to a regular weak value for  the energy
momentum tensor on the horizon, but we have not carried out an explicit
evaluation for such ``intermediate" cases. Of course the fact that the
weak energy can be either regular or singular on the horizon, depending
on the information available to observers with different state of
motion,   does not lead at this stage to any paradox but simply
emphasizes that detectable quanta on $\cal {I}^+$ require violent
energy-momentum fluctuations in the vicinity of the horizon and back to
$\cal {I}^-$.

This conclusion is not significantly altered when higher angular
momentum are included. The above description remains essentially valid
for all angular momenta except that most of the high angular momentum
modes are reflected back outside the star  towards the horizon.
Transplanckian fluctuations exist for all angular momenta but   outside
the shell the distribution of Hawking quanta is the result of a balance
of outgoing and reflected waves.  This brings them  nearly in thermal
equilibrium close to the horizon outside the star  with a local
blueshifted temperature  $$ T_{loc} = {1\over 8\pi M}(1-{2M\over
r})^{-1/2}
 \eqno(59)$$
 of the order of the blueshifted frequency there. Thus, the inclusion of
higher angular momentum for free fields does not change the
transplanckian character of the production process. Rather, it imbeds,
outside the shell, the s-wave transplanckian frequencies in a
transplanckian thermal bath Eq.(59) which would be unaffected by
interactions due do asymptotically free interacting renormalizable field
theory mixing different angular momenta. Conversely, the existence of
such a local temperature, given the potential barrier, is sufficient to
generate the Hawking radiation at infinity$^{[6],[8]}$.  \bigskip
\noindent {\bf 3. Weak values and Complementarity. } \medskip As
discussed in section 1, in presence of gravitational non-linearities,
frequencies  on  ${\cal I}^-$ are expected to be tamed to the Planck
scale by some hitherto unknown mechanism without altering significantly
the emitted radiation on  ${\cal I}^+$. In the shell model considered in
the preceding section,  transplanckian energies arose from quanta
travelling at  distances closer than the Planck size from  the horizon.
Within this toy model, these energies could be tamed to  Planck energies
if the planckian forces would bring the collapse to a halt at a
$\delta_2 = O(1/2M)$, or equivalently at a   Planck distance of its
Schwartzschild radius. This would constitute a dynamical taming of the
``hot" Boulware vacuum which emerged in section 2 from the description
in terms of the weak energy-momentum tensor from the post-selected
states available to external observers. Such an assumption is rather
natural if the background metric would be generated from weak values of
the gravitational field operators and not from their expectation
values;   these weak values   should   be sensitive to the weak values
of the energy-momentum source terms because, at least in the linear
approximation, weak values satisfy the same Heisenberg equations of
motion as the   expectation values.

 The assumption that, in presence of horizons, background are build from
weak values and hence dependent on post-selection will be discussed
below. We first examine how to give some substance to the toy model if
quantum gravity would indeed generate a background describing a stopped
collapse\foot{In a preliminary version$^{[21]}$, this scenario was
suggested but the conceptual framework discussed here was only
sketched.}. At the time $t_1$ given by Eq.(33) transplanckian
fluctuations appear inside the collapsing shell and at the time $t_2$
given by Eq.(34) they reach the Planck scale for an observer sitting
outside the shell at a fixed planckian distance from the horizon, that
is at $\delta_2 = O(1/2M)$, in a heat bath at the Planck temperature.
Thus the halting must take a time  $t^\prime_2 -t_1$ comparable or
larger than $t_2- t_1 = O(2M\ln 2M)$. During this time span the emission
process must be gradually altered and the original mechanism be
completely lost after the time $t^\prime_2 $ when the   background
geometry   becomes static. For the  radiation at ${\cal I}^+$ to survive
the halt of collapse, the local   temperature in the neighbourhood of
the shell must still be of the order of the Planck temperature.  But
such a thermal bath in the static geometry can only occur if   the
halting of the collapse has heated the collapsing matter, thereby
providing a source for the radiation, and such a heating is indeed
expected from energy conservation in a way similar to the heating of an
object maintained by external forces at a fixed radius from the horizon
of a black hole horizon. The role of the external force should be here
played by the unknown planckian forces responsible for the stopping.
However such a process  requires quantum transitions between the
constituents of the shell and the planckian thermal bath and cannot be
accounted for by the toy model.

 We therefore have   to  trade the classical shell and its surrounding
vacuum for a quantum  object  whose average classical configuration
tends towards a static one in the planckian neighbourhood of its
Schwartzschild radius. Such a picture seems to contradict Eqs.(46),(47)
and (48) which lead to zero flux for a static  classical trajectory.
However, these equations rely on the conservation law Eq.(45) which
should now be modified to take into account the source provided by
planckian constituents of the quantum shell. Phenomenologically one may
write $$  T_{\nu;\mu}^\mu(x)  = J_{\nu}(x) \eqno(60)$$  where
$J_{\nu}(x)$ represents the   source of the radiation arising from these
constituents    averaged over a Planck distance. In absence of a theory
of quantum gravity one may view this source term as an external one
and a possible term is, in the global $(u,v)$ system, $$\eqalignno{
J_u(u,v) &= 2 e^{-2\rho} k_u(u) \delta [v-v_s(u)] & (61)\cr J_v(u,v) &=
2 e^{-2\rho} k_v(v) \delta [u-u_s(v)] & (62)}$$ where $v_s(u)$ and
$u_s(v)$ are parametrizations of the shell averaged trajectory. The
functions $k_u(u)$ and $k_v(v)$   measure the radiation due the shell.
Instead of Eq.(46) and (47) we   get indeed outside the shell, adding
the source term to the trace anomaly,
 $$\eqalign {4\pi r^2   T_{uu}(x) &= {1\over 12\pi} \left[-{M\over
2r^3}(1-{2M\over r}) - {M^2\over 4r^4}  \right] + t_u(u) +   \Theta
[F(u,v)] k_u(u) \cr  4\pi r^2  T_{vv}(x)&= {1\over 12\pi} \left[-{M\over
2r^3}(1-{2M\over r}) - {M^2\over 4r^4}  \right] +   t_v(v)  - \Theta
[F(u,v)] k_v(v)}\eqno(63) $$ where $\Theta [F(u,v)]$ is a function which
goes through zero on the shell and is positive outside. The energy flux
across a sphere of radius $r$ at a retarded time $u$ becomes in general
 $$ T_{uu}^{(2)}- T_{vv}^{(2)}= t_u(u)+k_u(u). \eqno(64)$$ For   for
$u\simeq u_0$ it is all contained in $t_u(u)$ and gradually shifts to
the source  term $k_u(u)$ as the shell slows down and stops. One may
verify that in contradistinction with $t_u(u)$, $k_u(u)$ has
conventional tensor transformation under conformal reparametrizations.

The quantum shell represents only an outer region of finite thickness of
a realistic star.  Consider   a structureless classical shell of finite
thickness $\Delta L=\xi 2M$ at rest before collapse in the
asymptotically flat space time region. The shell may be viewed as part
of a larger star of mass $M$; $\xi$ is smaller than $1$ but still large
compared to $1/M$. At the classical level, the collapse of the shell
would induce a Lorentz contraction   to a coordinate separation $\Delta
r$ $$\Delta r = \left( 1-{2M \over r}\right)^{1\over2}
(1-v^2)^{1\over2}\Delta L \eqno(65)$$ where $v$ is the velocity of the
shell in the local static frame and hence $$ (1-v^2) =C \left(
1-{2M \over r}\right) \eqno(66)$$ where $C$ is of order $1$ and in fact
equal to unity   for
  a shell initially at rest at $r=\infty$. Thus for $t<t_0$, $\delta r
\simeq \xi \delta$, meaning that the   size   of the shell $\xi \sqrt
{(2M\delta)}$ shrinks to the Planck size between $t_1$ and $t_2$. To get
thermal equilibrium at the local Planck temperature, the shell should
stick to that size. For the classical shell this could occur neither if
the collapse went on (in which case its size would become too small) nor
if it would be brought to a halt (in which case the size would become
too big). But a quantum object composed of planckian constituents would
interact with the local Hawking excitations at that momentum scale. This
interaction could localize these constituents within a planckian
distance of the horizon, depending on the fundamental structure of
matter$^{[22]}$, and the whole shell could then remain localized on such
size.

Thus we envisage that the source of  Hawking radiation shifts gradually
from the vacuum fluctuations in the metric of a collapsing star to that
of a configuration which appears, on the average, static on time scales
short compared to $M^3$ that is to the evaporation time.   This means
that
  the background geometry has  a trivial topology.  Such
  structures where the star sticks within a Planck size of its
Schwartzschild radius have been introduced previously and labeled as
``achronons"$^{[14]}$.

For a realistic star, the process of terminating a collapse  into an
achronon heated at the Planck temperature   raises   a causality  issue.
The stopping of the star is a non local process and  halting the
perturbations around spherical symmetry requires correlations to be
established over the sphere in a time  comparable to the stopping time
which is   of the order of $4M \ln 2M$ or larger. These correlations
cannot be set in faster
 than the time necessary for a light ray  emitted during the collapse  to
reach the opposite point of the sphere. The shortest travel time for the
signal is achieved, when $\delta$ becomes of order $1$, by sending it
{\it outwards} to minimize time dilation effects. Outgoing  (and
ingoing) waves travelling radially in the neighbourghood of the horizon
take a time lapse $\Delta t$ to reach (to come from) a distance
comparable to the Schwartszchild radius  of order $$ \Delta t =2M \ln
{\delta_0 \over \delta(t)} \eqno (67) $$ where $t$ labels the emission
(arrival) time. Neglecting the time of order $M$ needed to travel on the
outer sphere  at $\delta = O(2M)$, we see that the travel time is of
order $\gamma M\ln 2M$ where $\gamma$ is a number between $4$ and $8$,
depending on the time of emission of the signal during the collapse
after the onset of Hawking radiation. This seems to indicate that there
is hope that such correlations can be established\foot{it was argued in
reference $[22]$ that strings could establish   correlations over the
horizon in the time spent by a signal emitted at a Planck distance from
the horizon, that is for $\gamma=8$.}.

The mechanism considered above to get rid of transplanckian frequencies
from planckian effects would keep the source of Hawking radiation in
causal contact with the  star and remain so during its subsequent
evaporation. After a short ignition period during which the collapsing
star emits thermal quanta uncorrelated to its structure, the achronon
burns away  its mass and structure as any hot body does within a
topologically  trivial background  void of horizons and of
singularities. Although we have not implemented our description by
dynamics, we can  see no reason why the   evaporation process should not
be viewed as a unitary process within this background at least to the
extend that background transitions are small.

This is at odds with the usual picture assuming weak back-reactions
effects induced by the average energy-momentum tensor of the radiation.
In this case, except  at the final stage, one could treat in good
approximation the emission process adiabatically by fixing the classical
Schwartzschild background by the instantaneous mass. The cumulative
effect of the varying mass leads however to qualitative
effects$^{[23],[24]}$ which we briefly discuss. For a constant mass the
global event horizon coincided  outside the star with the boundary of
trapped surfaces, that is with the apparent horizon. The latter now
separates from the former.  A
hypothetical light ray which would be emitted from inside the star and
cross the shell between the two horizons would first recede to smaller
radius and diverge again upon reaching the apparent horizon. The
evaporation of the black hole is not due to a reduction of the shell
mass but results from a negative contribution to the total mass from a
cloud concentrated between the event and the apparent horizon.
 This cloud is then situated at
smaller  radius than   the shell when the latter was in causal contact
with the distant observer but is separated by a space-like distance from
the shell at the same radius. Thus the march of the classical star
towards its final destruction  by tidal forces in the vicinity of the
classical singularity appears as a causally disconnected history from
the evaporation out of a polarisation cloud which fills up with
increasing mass (in absolute value)  a macroscopic volume of space
``outside the star surface" back to $r\simeq 0$.

This  sequence of events is fully consistent with  a thermal,
structureless Hawking radiation encoding no information about the
original star. All the information is contained in the collapsing star
evolving with its initial mass. This is also consistent with the fact
that, as required by the quantum superposition principle, there is no
duplication of information between the collapsing star and the Hawking
radiation$^{[6]}$. In fact, the latter argument seem to indicate that
the picture emerging from the semi-classical back-reaction anzats is
bound to survive quantum corrections and that up to the Planck mass
scale, the Hawking radiation of a collapsing star cannot, even in
principle, contain any relevant information about the detailed structure
of the star. One would then be confronted with the usual dilemma of
either a full evaporation and the concomitant loss of unitarity  in the
universe left behind or with the halting of the collapse at the external
observers Planck size, relegating there the huge and presumably infinite
degeneracy of a left over remnant.

The termination of a collapse into an achronon would cure
tranplanckian effects and possibly the unitarity issue, but appears in
blatant contradiction with the equivalence principle as there can be no
classical force to generate the huge acceleration needed to halt the
collapse in a region where a free-falling object, would, according to
classical general relativity feel, for large collapsing masses,
vanishingly small curvature. A reconciliation was already suggested
above from the presumed dynamical origin of the taming. The tamed metric
background in which the black hole    evaporates would stem from   to
weak values of the energy-momentum tensor; it would not coincide with
the expectation value of $g_{\mu\nu}$ in the Heisenberg state $\ket i$
defined by the star prior to collapse but rather by some average weak
value $<g_{\mu\nu}(x)>_{0}$. This value would be the result of an
averaging over post-selected states defined by a complete set of
commuting  observables describing, on a future space like surface,  the
Hawking quanta due to the complete evaporation to  zero final mass of
the hole. It would
 describe a geometrical background with a trivial achronon topology
  in which the future
space-like surface where post-selection of Hawking quanta is performed
is a Cauchy surface. In contradistinction
with this reconstructed history of the collapse by external observers,
the proper history of the star, as recorded by a comoving observer,
follows from the back-reaction to a trivial post-selection corresponding
to essentially no detectable Hawking quanta. This back-reaction should
  not destroy the classical background and   the star would cross the
horizon and end up in the singularity, in accordance with the equivalence
principle. This background we label as   $<g_{\mu\nu}(x)>_{M}$ to recall
that the star retains its full mass M.   One must then also consider
intermediate geometries detected by observers recording Hawking
radiation up to some final  mass $m$, $0<m<M$ and then collapsing in the
background of a star of mass $m$. Thus we could classify the histories
of the star according to classes defined by some classical parameter,
say $m$, labeling as in Eq.(9) averages of weak values over final states
with the corresponding parameters: $$ <g_{\mu\nu}(x)>_{m}=
\sum_{{f_m}\epsilon m}P_{f_m} o  {\bra {f_m}\hat g_{\mu\nu}(x)
\ket i \over \bra {f_m} i\rangle}; \qquad P_{f_m}\equiv {\vert\bra {f_m}
i\rangle\vert^2 \over \sum_{f'_{m'}} \vert\bra {f'_{m'}}
i\rangle\vert^2}\eqno(68)$$ Thus in presence of the classical event
horizon an infinity of different background geometries would be
generated according to the post-selected events available to a given
observer. This form of complementarity is somewhat different from the
one proposed in reference [6] where two histories were assumed: the one
available to external observers and the proper one. Consistency for
observers crossing the horizon after partial evaporation of the hole is
then tentatively ensured by arguing that these observers cannot receive
cisplanckian signals from the proper motion before reaching the
singularity$^{[25]}$. The present approach  based on weak values
associate different geometries and hence different histories to
detectable differences in the amount of Hawking radiation recordable
before crossing the horizon.  Histories are   relative to the
post-selections  available to the state of motion of the observer and
split upon the latter crossing the horizon\foot{The observer dependence
of effects due to back-reaction in presence of horizons was suggested
previously   by Gibbons and Hawking$^{[12]}$.}. In each history,
unitarity is expected to be satisfied to the extend that transitions
between different background can be neglected. To avoid paradoxes in
this enlarged concept of complementarity,  interferences between
macroscopically distinct histories must be negligible, as is generally
the case in ordinary quantum physics for macroscopic systems containing
many degrees of freedom. A more dramatic possibility would be that
quantum gravity allow fluctuations of the geometry only in the vicinity
of a classical background. This alternative would attach to each
configuration  a well defined meaning to time as the latter is an
operational concept only in the vicinity of a classical
geometry$^{[26]}$. Finally we want to   stress that the fact that weak
values are called upon for selecting backgrounds does not conflict with
causality. Causality is not posed in different terms for weak values
than for the more conventional expectation value ansatz: one just
selects different matrix elements of the same Heisenberg operator.

Clearly, in absence of a reliable theory of quantum gravity, Eq.(68) has
only formal and provisional meaning. In particular, it is not clear how
to cope with a possibly complex background metric. Also the very notion
of a background geometry can only be made precise if physics admits a
description in terms of objects defined on it. Despite and perhaps
because of these ambiguities, the possible relevance of our formulation
of complementarity is to provide a germ of a possible conceptual
framework in which the dynamics of quantum gravity should   fit. At
present all one can say is that, as mentioned above, the arguments of
reference [22] seem to indicate, in a different but related context,
that elements of string theories may be relevant in particular for
ensuring the Lorentz transformation properties of matter constituents
and the correlations over the whole horizon needed to reconstruct the
star history by the external observers. In addition the taming of
transplanckian frequencies outside the star follows also rather
naturally from the spectrum of string theories$^{[8]}$. Thus string
theories may contain some of the ingredients required   to realize
the above scheme which appears  clearly inconsistent with the
conventional local field theory whose shortcomings motivated its
construction. \vfill\eject \centerline {\bf Acknowledgements} \medskip I
am particularly grateful to Y. Aharonov for his help in clarifying   the
conceptual issues involved in this work. I also thank R. Argurio, R.
Brout, A. Casher, S. Massar,  R. Parentani and Ph. Spindel for useful
discussions  \vfill\eject

\centerline{REFERENCES}

\item{[1]} S.W. Hawking, Commun. Math. Phys. {\bf 43} (1975) 199.

\item{[2]} S.W. Hawking, Phys. Rev. {\bf D14} (1976) 2460 .

\item{[3]} Y. Aharonov, A. Casher and S. Nussinov, Phys. Lett. {\bf
B191} (1987) 51.

\item{[4]} G.'t Hooft, Nucl. Phys. {\bf B335} (1990) 138

\item{[5]}  G.'t Hooft, in the Proceedings of the ``International
Conference on Fundamental Aspects of Quantum Theory" (1992), in Honour
of Y. Aharonov's 60th birthday.\hfill \break C.R. Stephens, G.'t Hooft
and B.F. Whiting, ``{\it Black Hole Evaporation
 without Information Loss}", Preprint THU-93/20,
UF-RAP-93-11,gr-qc/9310006 (1993).

\item{[6]} L. Susskind, L. Thorlacius and J. Uglum, Phys. Rev. {\bf
D48}(1993) 3743.

\item{[7]} S. Massar and R. Parentani, ``{\it From Vacuum Fluctuations
to Radiation:
 Accelerated Detectors and Black Holes}" Preprint ULB-TH 94/02,
gr-qc/9404057 (1994).

\item{[8]} F. Englert, S. Massar and R. Parentani  Class. Quantum Grav.
{\bf 11} (1994) 2919.

\item{[9]} J.D. Beckenstein, Phys. Rev. {\bf D7} (1973)  2333.

\item{[10]} J.M. Bardeen, B. Carter and S.W. Hawking, Comm. Math. Phys.
{\bf 31} (1973) 161.

\item{[11]} A. Casher and F. Englert, Class. Quantum Grav. {\bf 9}
(1992) 2231.

\item{[12]} G. Gibbons and S. Hawking, Phys. Rev. {\bf D15} (1977) 2738.

\item{[13]} G. Gibbons and S. Hawking, Phys. Rev. {\bf D15} (1977) 2752.

\item{[14]} A. Casher and F. Englert, Class. Quantum Grav. {\bf 10}
(1993) 2479.

\item{[15]} G.'t Hooft, J. Geometry and Phys. {\bf 1} (1984) 45.

\item{[16]} Y. Aharonov, D. Albert, A. Casher and L. Vaidman, Phys.
Lett. {\bf A124} (1987) 199. \hfill \break Y. Aharonov and L. Vaidman,
Phys. Rev. {\bf A41} (1990) 11. \hfill \break Y. Aharonov, J. Anandan,
S. Popescu and L. Vaidman, Phys. Rev. Lett. {\bf 64} (1990) 2965.

\item{[17]} W.G. Unruh, Phys. Rev. {\bf D14} (1976) 287.

\item{[18]} R. Parentani and R. Brout, Int. J. Mod. Phys. {\bf D1}
(1992) 169.

\item{[19]} P.C.W. Davies, S.A. Fulling and W.G. Unruh, Phys. Rev. {\bf
D13} (1976) 2720.\hfill

\item{[20]} R. Parentani, Class. Quantum Grav. {\bf 10} (1993) 1409.

\item{[21]} F. Englert ``{\it The Black Hole History in Tamed Vacuum}",
Preprint ULB-TH 15/94,
 gr-qc/9408005 (1994).

\item{[22]} L. Susskind ``{\it The World as a Hologram}", Preprint
SU-ITP-94-33,
 hep-th/9409089 (1994).

\item{[23]} J.M. Bardeen, Phys. Rev. Letters {\bf 46} (1981) 382.

\item{[24]} R. Parentani and T.Piran, ``{\it The Internal Geometry of an
Evaporating Black Hole}" hep-th/9405007 (1994).

\item{[25]} L. Susskind and L. Thorlacius, Phys. Rev. {\bf D49} (1994)
966.

\item{[26]} T. Banks, Nucl. Phys. {\bf 249} (1985) 332.\hfill\break  R.
Brout, Foundations of Physics {\bf 17} (1987) 603

\end